\documentclass[a4paper,11pt]{article}
\pdfoutput=1
\usepackage{jinstpub}
\usepackage{graphicx} 
\usepackage{url}

\newcommand{\Itot}{$I_{\mathrm{tot}}$}
\newcommand{\Ig}  {$I_{\mathrm{G3D}}$}
\newcommand{\Ime}  {$I_{\mathrm{mesh}}$}
\newcommand{\Ea}  {$E_{\mathrm{A}}$}
\newcommand{\La}  {$L$(\Ea)}
\newcommand{\nel}  {$n_{\mathrm{el}}$}
\newcommand{\ngem}  {$n_{\mathrm{GEM}}$}
\newcommand{\Vg}  {V$_{\mathrm{GEM}}$}
\newcommand{\agem}  {$\alpha_{\mathrm{GEM}}$}

\newcommand{\nm}  {$n_{\mathrm{exc}}$}
\newcommand{\aex}  {$\alpha_{\mathrm{exc}}$}

\title{\boldmath First evidence of luminescence in a He/CF$_4$ gas mixture induced by non-ionizing electrons}

\author[a,b]{E. Baracchini,} 
\author[c]{L. Benussi,}
\author[c]{S. Bianco,}
\author[c]{C. Capoccia,} 
\author[c,d]{M. Caponero,}
\author[e,f]{G. Cavoto,}
\author[a,b]{A. Cortez,}
\author[e]{I. A. Costa,}
\author[e]{E. Di Marco,}
\author[e]{G. D'Imperio,}
\author[a,b]{G. Dho,}
\author[e]{F. Iacoangeli,}
\author[c]{G. Maccarrone,}
\author[e,g]{M. Marafini,}
\author[c]{G. Mazzitelli,}
\author[e,f]{A. Messina,}
\author[c]{A. Orlandi,}
\author[c]{E. Paoletti,}
\author[c]{L. Passamonti,}
\author[h,i]{F. Petrucci,}
\author[c]{D. Piccolo,}
\author[c]{D. Pierluigi,}
\author[e,1]{D. Pinci,\note{Corresponding author.}}
\author[e]{F. Renga,}
\author[c]{F. Rosatelli,}
\author[c]{A. Russo,}
\author[c,j]{G. Saviano,}
\author[c]{and S. Tomassini}

\affiliation[a]{Gran~Sasso~Science~Institute,\\ L'Aquila, I-67100, Italy}
\affiliation[b]{Istituto Nazionale di Fisica Nucleare,\\ Laboratori Nazionali del Gran Sasso, Assergi, Italy}
\affiliation[c]{Istituto Nazionale di Fisica Nucleare ,\\  Laboratori Nazionali di Frascati, I-00044, Italy}
\affiliation[d]{ENEA Centro Ricerche Frascati, Frascati, Italy}
\affiliation[e]{Istituto~Nazionale~di~Fisica~Nucleare,\\ Sezione di Roma, I-00185, Italy}
\affiliation[f]{Dipartimento di Fisica Sapienza Universit\`a di Roma, I-00185, Italy} 
\affiliation[g]{Museo Storico della Fisica e Centro Studi e Ricerche "Enrico Fermi",\\ Piazza del Viminale 1, Roma, I-00184, Italy}
\affiliation[h]{Dipartimento di Matematica e Fisica, Universit\`a Roma TRE, Roma, Italy}
\affiliation[i]{Istituto Nazionale di Fisica Nucleare, Sezione di Roma TRE, Roma, Italy}
\affiliation[j]{Dipartimento di Ingegneria Chimica, Materiali e Ambiente, Sapienza Universit\`a di Roma, Roma, Italy}

\emailAdd{davide.pinci@roma1.infn.it}

\abstract{

Optical readout of Gas Electron Multipliers (GEM) provides very interesting performance and has been proposed for different applications in particle physics. In particular, thanks to its good efficiency in the keV energy range, it is being developed for low-energy and rare event studies, such as Dark Matter searches.
So far, the optical approach has only exploited the light produced during the avalanche processes in GEM channels.

Further luminescence in the gas can be induced by electrons accelerated by a suitable electric field.
The CYGNO collaboration studied this process with a combined use of a triple-GEM structure and a grid in an He/CF$_4$ (60/40) gas mixture at atmospheric pressure. 
Results reported in this paper allow to conclude that with an electric field of about 11~kV/cm a photon production mean free path of about 1.0~cm was found.}

\begin{document}
\maketitle
\flushbottom

\section*{Introduction}

The reading-out of the light produced by the de-excitation of gas molecules during the processes of electron multiplication in Micro Pattern Gaseous Detectors (MPGDs) has shown a very fast development in recent years~\cite{bib:ref1,bib:ref2,bib:ref3,bib:nim_orange1,bib:jinst_orange1}.

This approach has become possible thanks to the significant progress achieved both in the performance of MPGDs as well as in the evolution of the light sensor technology. Complementary Metal Oxide Semiconductor (CMOS) based devices combine high sensitivity with a low noise level and provide very good performance for detecting and tracking particles in the gas~\cite{bib:nim_orange2,bib:eps, bib:jinst_orange2}. 

The granularity of such CMOS-based devices requires a good production of photons to properly illuminate its large number of pixels. Therefore, the total amount of light produced is still a dominant limiting factor.

A possibility of increasing the light signal is given by the exploitation of the electron-induced luminescence in the gas outside the multiplication regions.

In this paper the result of measurements of electo-luminescence in He/CF$_4$ (60/40) gas mixture at atmospheric pressure is reported.

The R\&D presented is part of the CYGNO work to develop an Optical Readout based TPC for Dark Matter searches and rare processes study.

\section{Electro-luminescence in He/CF$_4$ based mixtures}
\label{sec:he-cf4}
Light produced in a He/CF$_4$ gas mixture by primary and secondary ionization processes was widely studied in the past \cite{bib:Fraga,bib:Margato2}.
The conclusions of these studies that have important impacts on this paper, can be summarised here:

\begin{enumerate}
    \item The emission spectrum of He/CF$_4$ (60/40) exhibits two broad molecular continua: one centred at 290~nm with a shoulder at 240~nm and another centred at 620~nm which extends from 500~nm to 750~nm;
    \item The broad band in the visible region results from the excitation of a Rydberg state of the CF$_4$ molecule that dissociates into an emitting CF$^*_3$ fragment. The energy threshold for this emission is found to be 12.5~eV;
    \item This threshold is not far from the value of 15.9~eV of the dissociative ionisation threshold;
    \item A number of about 3 collisions per centimetre (i.e. a mean free path of 3.3~mm) leading to dissociation of CF$_4$ into neutral fragments, was evaluated for an electric field of 12.5~kV/cm~\cite{bib:Fraga}. This value rapidly decreases for lower electric field values and is expected to be around 1 collision per centimetre at 11.0~kV/cm.
\end{enumerate}

\section{Experimental setup}

\subsection{The detector}

\label{sec:setup}
\begin{figure}[ht]
\centering
\includegraphics[width=0.9\textwidth]{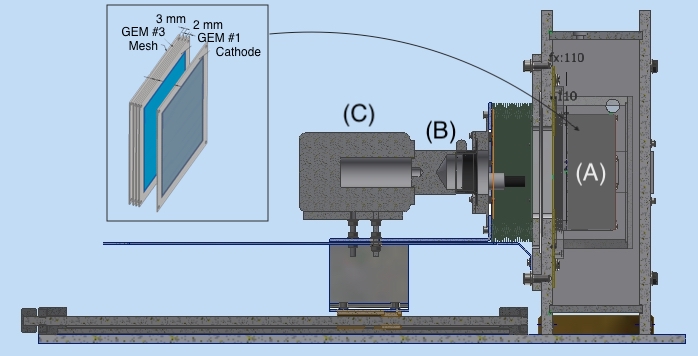}
\caption{Schematic of the experimental setup. The position of sensitive volume is indicated (A) and the lens (B) and the CMOS camera (C) are visible.
In the inset, the triple-GEM structure with the mesh and cathode is visible.} 
\label{fig:setup}
\end{figure}

Figure~\ref{fig:setup} shows a sketch of CYGNO prototype used in these measurements.
Three $10\times10$~cm$^2$ standard GEMs (50~$\mu$m thickness, 70~$\mu$m diameter holes with 140~$\mu$m pitch)
are mounted with two 2~mm wide transfer gaps between them.
A planar cathode is placed in front of GEM~\#1 to create a 1~cm wide drift gap that represents the sensitive volume of the device and that results therefore to be 100~cm$^3$.

Beyond the third GEM (GEM~\#3), at a distance $\Delta z$ of 3 mm, a metal mesh is placed to provide an electric field to accelerate electrons exiting from the GEM channels. Its optical transparency was estimated to be around 60\%.

In this configuration, the currents drawn by the bottom electrode of GEM~\#3 and by the mesh were both measured by a current-meter with a sensitivity of 2~nA.

A transparent plastic foil
is used to seal the gas volume beyond the third GEM.
With a transmission larger than 90\% for visible light, 
it allows the detector to be optically coupled to the readout system, 
placed outside the detector gas volume.
An Hamamatsu Orca Flash 4.0 camera, based on a CMOS sensor, 
is used\footnote{For more details visit http://camera.hamamatsu.com/eu/en/product/search/C11440-42U/index.html}
to acquire images produced by the photons created in the gas ((C) in Fig.\ref{fig:setup})
through an 8~cm side squared window. 

As described in \cite{bib:jinst_orange1},
the sensor ($2048 \times 2048$ pixels) 
has a noise level of less than 2 electrons per pixel,
a quantum efficiency of about 70\% at 600 nm
and provides 0.9 counts per photon.
The camera is equipped with a Schneider 
lens\footnote{For more details visit https://www.edmundoptics.com/imaging-lenses/fixed-focal-length-lenses/Schneider-Fast-C-Mount-Lenses/} 
providing an aperture $a = 0.95$ with a focal length of 25 mm ((B) in Fig.\ref{fig:setup}).
The lens is placed at a distance $d$ of 27.5 cm from the last GEM
in order to obtain a de-magnification 
$\delta = (d/f) - 1 = 10$ to 
image a $13.3 \times 13.3$~cm$^2$ surface onto the 
$1.33 \times 1.33$~cm$^2$ sensor. 
In this configuration, each pixel
looks at an effective area of 
$65 \times 65$~$\mu$m$^2$.
The fraction of the light collected by the lens can be evaluated to be $5.7 \times 10^{-4}$~\cite{bib:ieee_orange}.

\subsection{Data acquisition and analysis}

All data described in this paper refers to runs where gas in the drift gap was exposed to the 5.9~keV photons provided by a $^{55}$Fe source placed just outside the sensitive volume corresponding to the space between the cathode and GEM~\#1.
The nominal source activity was 740~MBq.

Given the very high activity of the source and the small size of the sensitive volume, the rate of signals due to natural radioactivity was measured to be more than ten times lower than the one due to 5.9~keV photons. For this reason it was possible to operate the camera acquisition in {\it free running} mode (images are continuously acquired and stored).

As described in \cite{bib:fe55}, the interaction of low energy photons in the gas mixture in use create photo-electrons that release their whole energy in few hundreds of micrometers. Because of subsequent diffusion of electrons during the drift toward the GEM, these events produce round spot-like signals on the CMOS sensor with diameters of 2-3~mm.

The detector was operated in a constant configuration: 
\begin{itemize}
    \item the electric field in the drift gap was kept to 0.9~kV/cm;
    \item the electric field in the gaps between the GEM equal to 2.5~kV/cm;
    \item voltage difference across the GEM (\Vg) electrodes was set to 400~V;
    \item flushed with an He/CF$_4$ (60/40) mixture at a total rate of 100~cc/min at atmospheric pressure.
\end{itemize}

In this configuration, according to previous measurements \cite{bib:ieee_orange}, the triple-GEM structure is expected to provide a total charge gain of the order of 10$^5$.
While keeping the electrical configuration of the stack stable, the voltage difference between the bottom electrode of GEM~\#3 and the mesh ($\Delta$V) was scanned from 0~kV to 3.4~kV. This upper value was chosen because of some electrical instabilities arising for a $\Delta$V of 3.5~kV.
Therefore, the electric field in the acceleration gap (\Ea) was scanned in the range [0-11.3]~kV/cm. 

For each value of $\Delta$V two sets of data were taken:

\begin{itemize}
    \item {\it Long}: 30 images with a sensor exposure of 10 seconds. 
    \item {\it Short}: 200 images with a sensor exposure of 500 ms. 
\end{itemize}

The difference in the amount of collected light and spot occupancy in the images in the two run types allow to extract similar information with different analysis methods in order to cross-check the results.

For the {\it Long} runs a very simple algorithm was developed with no spot identification or reconstruction: the total amount of light reaching the sensor \La\ is simply evaluated as the average over all images in the run of the sum the number of photons detected by each pixel. The RMS of these sums is assumed as its uncertainty.

On the other hand, the images in the {\it Short} runs, were analysed to reconstruct the number of photons collected in each single spot. Spot reconstruction and analysis was performed with a {\it nearest neighbor clustering} algorithm and the obtained distributions were fitted with a Gaussian plus a Polya function as described in \cite{bib:fe55}. This analysis allows to study the behavior of the signal and separate it from possible variation of light background.

\section{Experimental results and discussion}

\subsection{Light yield behaviour}
\label{sec:ly}
\begin{figure}[ht]
\centering
\includegraphics[width=0.8\textwidth]{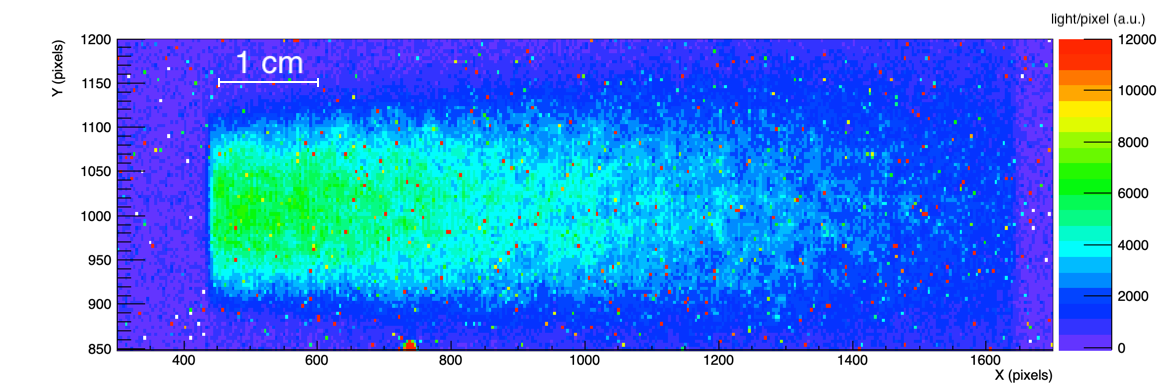}\\
\includegraphics[width=0.8\textwidth]{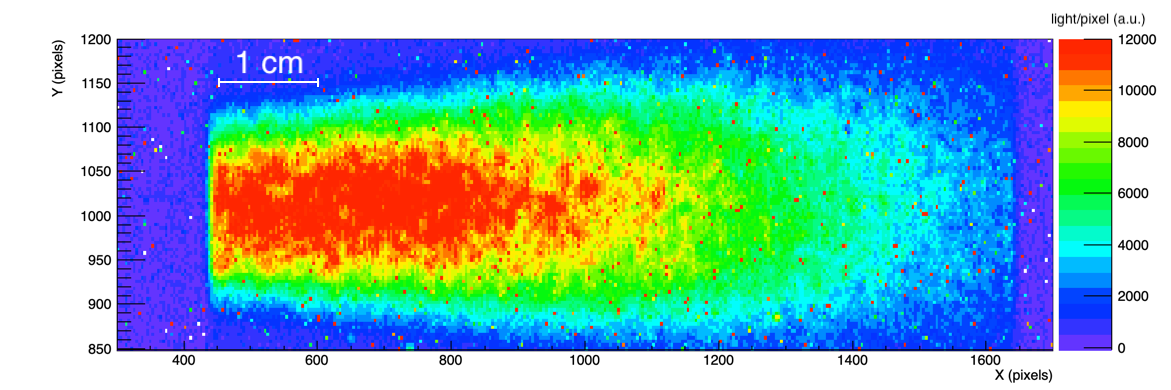}
\caption{Map of the light detected in the detector for \Ea~=~0~kV/cm (top) and for \Ea~=~11.3~kV/cm (bottom).} 
\label{fig:beam}
\end{figure}

Figure \ref{fig:beam} shows the spatial distribution of the recorded events for \Ea~=~0~kV/cm (top) and for \Ea~=~11.3~kV/cm (bottom) in the central part of active area (the borders of the
8~cm wide transparent windows are visible for $x$ around 430~pixels and 1650~pixels).
These two plots show data from two {\it Long} runs.

\begin{figure}[h!]
\centering
\includegraphics[width=0.80\textwidth]{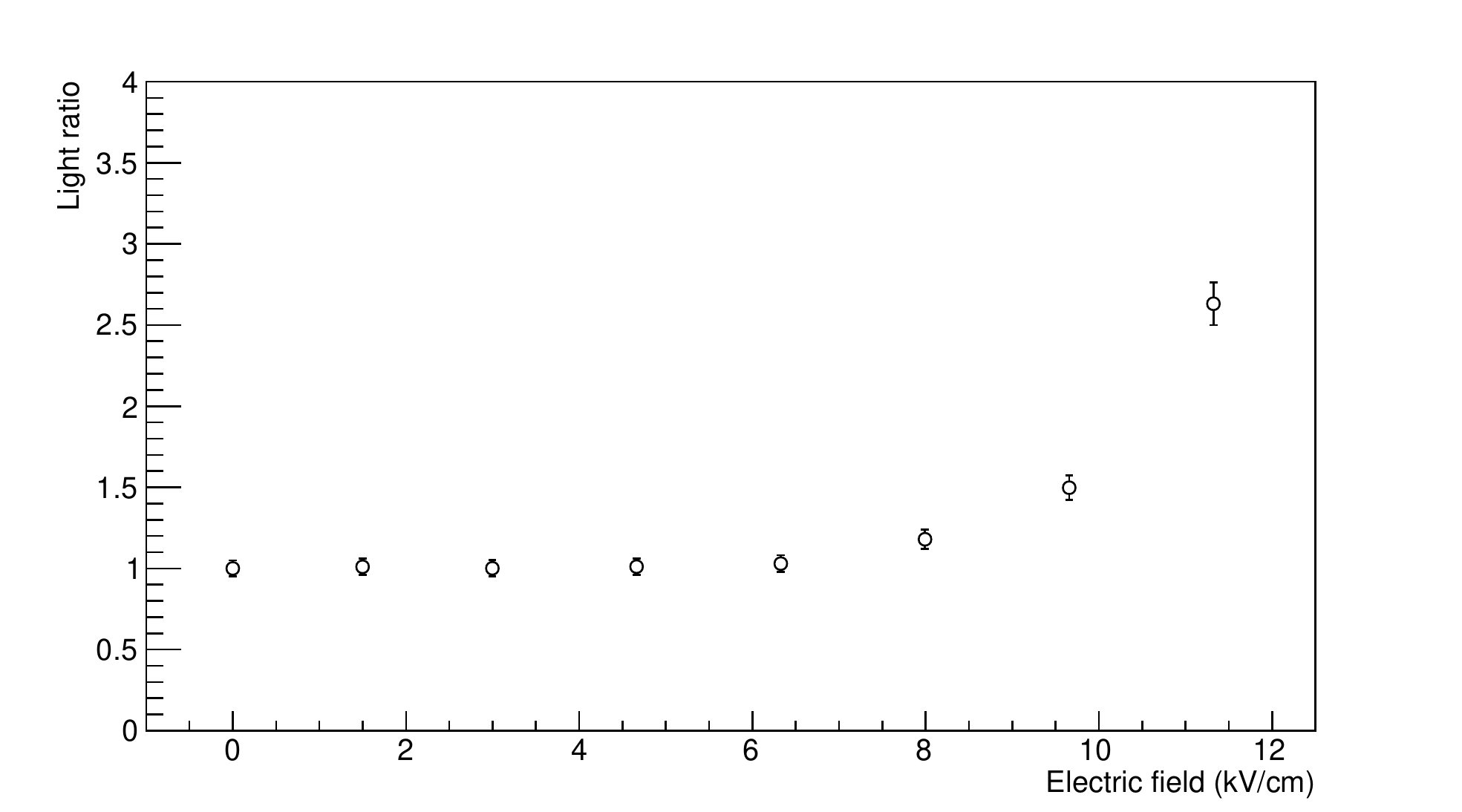}
\caption{Behavior of total amount of light ratio ($L$(\Ea)/$L$(0)) as a function of \Ea} 
\label{fig:light}
\end{figure}

The acceleration gap is clearly effective in increasing the light detected.
For each value of \Ea, \La\ is evaluated.
The behavior of $L$(\Ea)/$L$(0) is shown in Fig.\ref{fig:light} as a function of \Ea.

It is possible to see that, up to values of \Ea\ of about 6.0 kV/cm \La\ is constant and, therefore, the only light collected is the one produced in GEM amplification processes. Starting from 
\Ea~=~8.0~kV/cm, the light produced rapidly increases and a light yield 2.6 times larger is recorded for \Ea~=~11.3~kV/cm.

\subsection{$^{55}$Fe spectra studies}
\label{sec:55fe}
\begin{figure}[h!]
\centering
\includegraphics[width=0.407\textwidth]{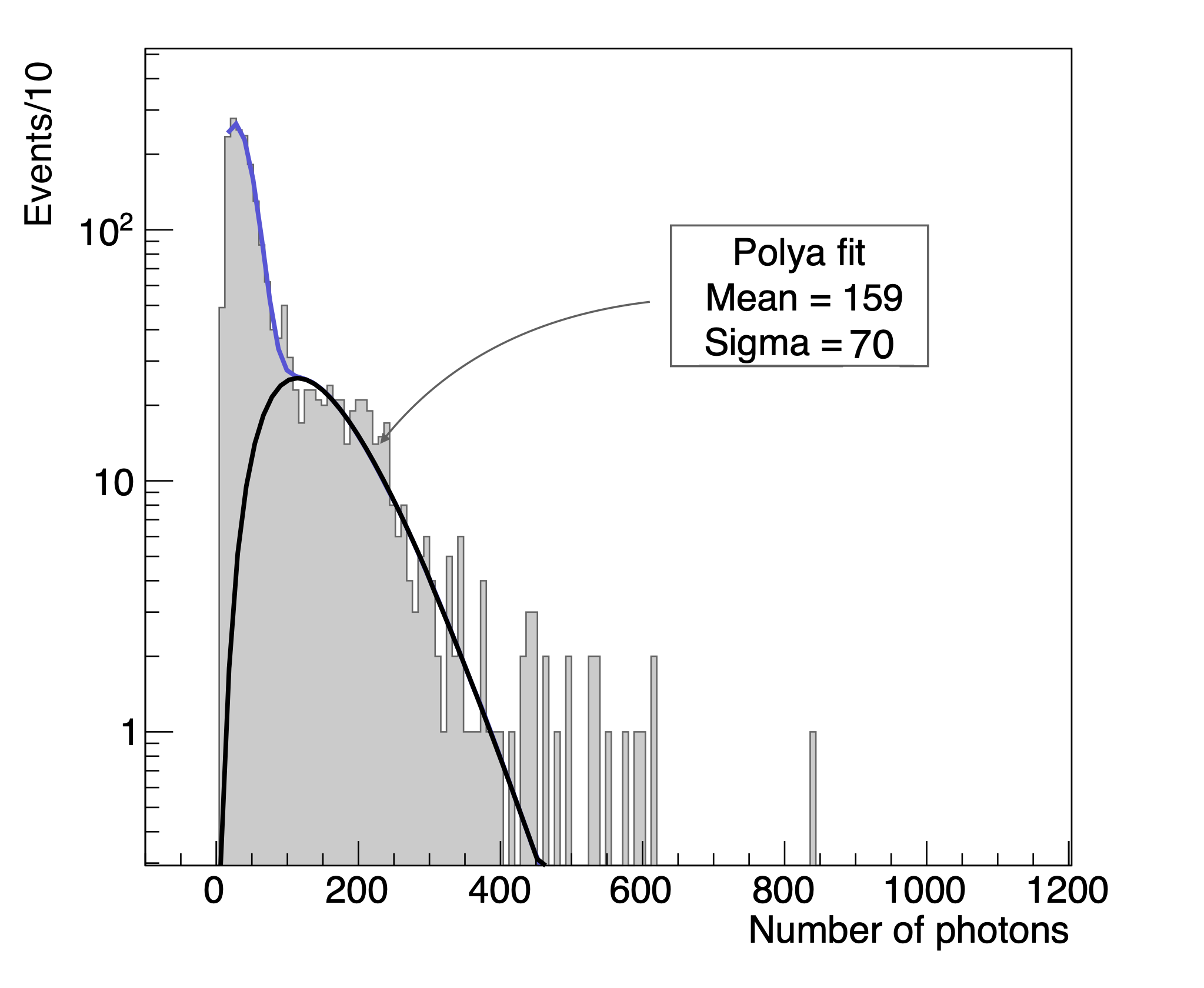}
\includegraphics[width=0.4\textwidth]{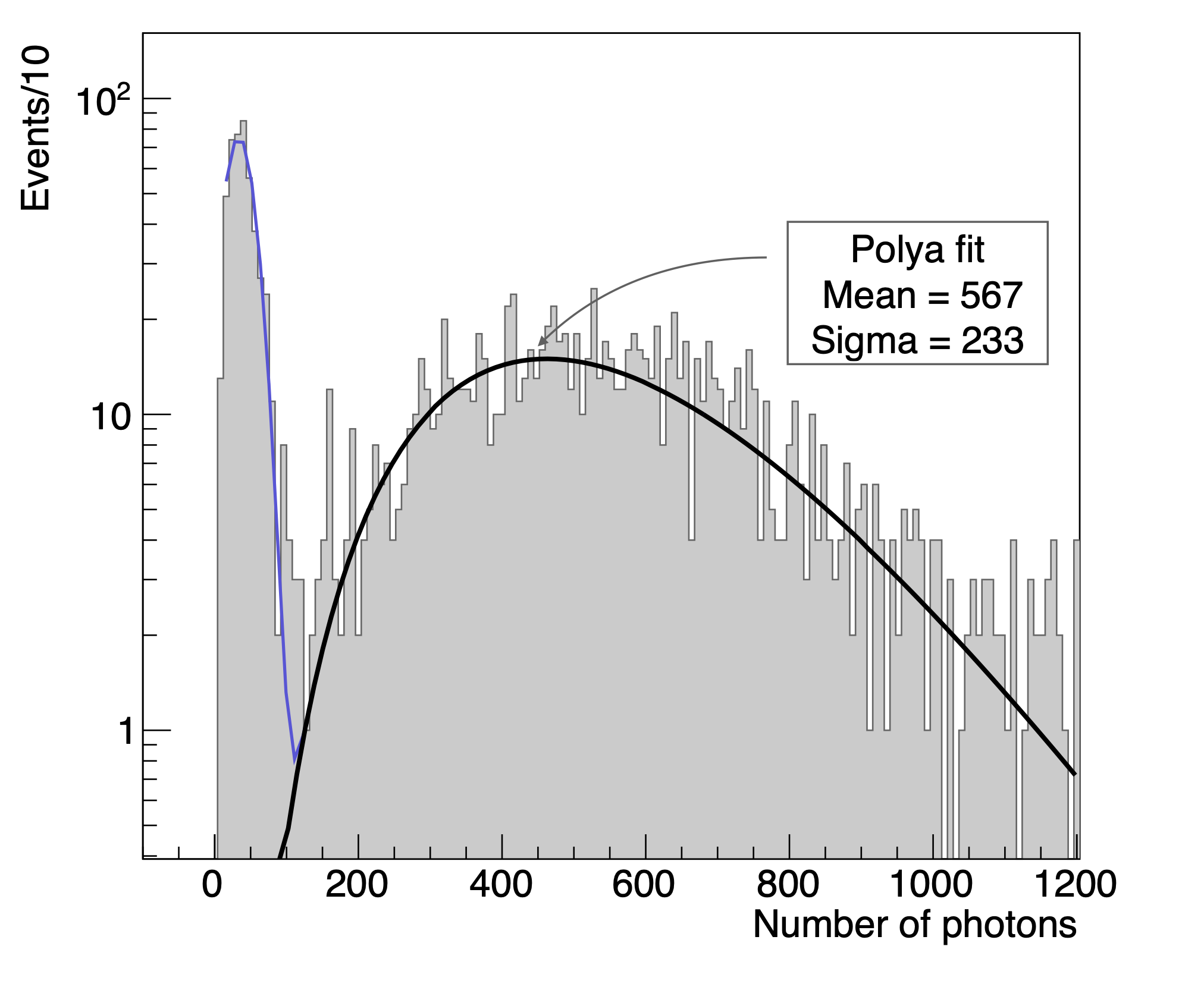}
\caption{$^{55}$Fe spectra produced in the detector for \Ea~=~0~kV/cm (left) and for \Ea~=~11.3~kV/cm (right), with a fit to a Polya for the signal plus a Gaussian for the noise superimposed in blue, and only the Polya component superimposed
in black.} 
\label{fig:qIn}
\end{figure}

Two examples of distribution of light in $^{55}$Fe reconstructed spots are shown in Fig.~\ref{fig:qIn} with  fits to a Polya for the signal peak plus a Gaussian for the noise superimposed in blue and the only Polya component superimposed in black (for details see \cite{bib:fe55}).

Given the high number of collected signals, the uncertainties in the extracted fit parameters 
were always found to be below of 1-2\%. 
Reconstructed pedestal position was stable within less than 10\% in the various runs and, thanks to the very low noise level of the sensor, its fluctuations were measured to be about 25 photons, always well below the signal ones.
The large increase in the light associated to each $^{55}$Fe spot for high values of \Ea\ is well visible. 
It should also be noticed that, the increased signal does not spoil the relative energy resolution of the detector (around 40\% in both cases because of the low \Vg\ value was chosen). 

\begin{figure}[ht]
\centering
\includegraphics[width=0.80\textwidth]{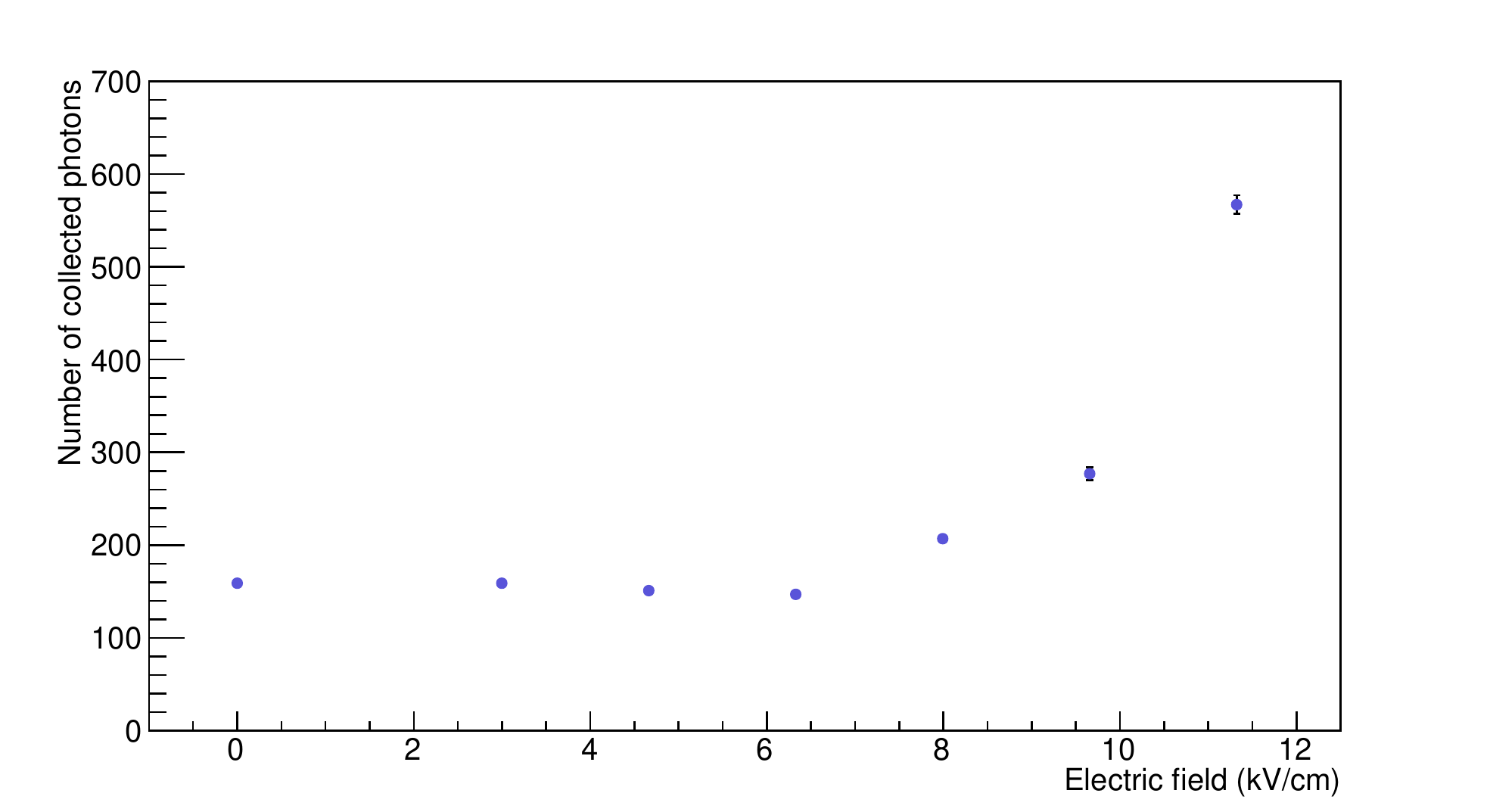}
\caption{Behaviour of the fitted mean value of the Polya function for the
$^{55}$Fe spectra as a 
function of \Ea.} 
\label{fig:qIn2}
\end{figure}

The behavior of the peak position as a function of \Ea\ is shown in Fig.~\ref{fig:qIn2}. 
The associated uncertainties are given by the Polya fit standard deviation normalised 
with the square root of the number of reconstructed spots.

The trend of spectra mean values confirms the result obtained in the \La\ analysis (see Sect.~\ref{sec:ly}): almost constant for \Ea\ lower than 8~kV/cm and a rapid increase for large \Ea\ values.

\subsection{Charge studies}

During the whole data taking, the currents drawn by the bottom electrode of last GEM (\Ig) and by the mesh (\Ime) were continuously monitored. 

The values of the two currents were first measured in absence of the $^{55}$Fe source for different \Ea\ in the range [0-11.3]~kV/cm values and found to be negligible and compatible with the current meter sensitivity.

\begin{figure}[ht]
\centering
\includegraphics[width=0.80\textwidth]{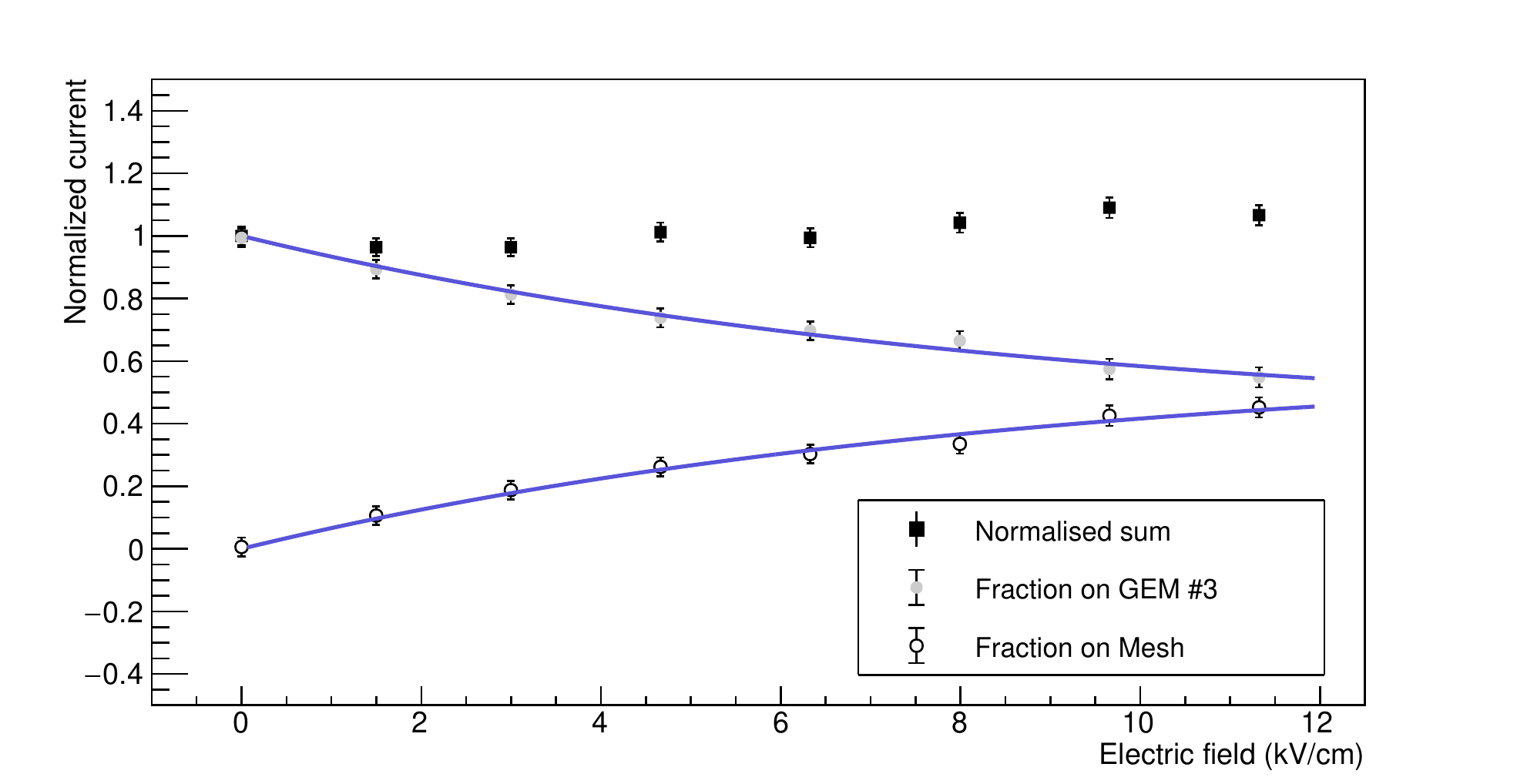}
\caption{Sum of \Ig\ and \Ime\ normalised to \Ea~=~0~kV/cm (see text for details) as a function of \Ea\  (black circles) together with their 
fraction with respect to the sum.} 
\label{fig:currents}
\end{figure}

The behavior of the sum \Itot=\Ig +\Ime\ in presence of the $^{55}$Fe, as a function of 
\Ea, normalised to its initial value is shown in Fig.~\ref{fig:currents}.

Except for a small increase for \Ea\ values larger than 8~kV/cm, \Itot\ was quite stable within less than 15\% and almost independent from \Ea. This result indicates the substantial stability of the total charge in the presence of various voltage differences between the GEM~\#3 and the mesh. An increase of this quantity would have indicated a variation in the multiplication processes within the GEM~\#3 or the arise of avalanches on the mesh wires.
This stability allows, instead, to conclude that further light detected by the camera is produced by simple electro-luminescence of the gas in the gap between GEM~\#3 and the mesh, induced by electrons exiting from GEM channels, without any detectable further gas ionisation.

Moreover, the behavior of ratios  \Ig/\Itot\ and \Ime/\Itot\ provides information about the charge sharing between the two electrodes.
For \Ea~=~0 kV/cm all charge exiting from multiplication channels of GEM~\#3 is collected by the bottom electrode of the GEM. In presence of an electric field below the GEM electrons are extracted and drifted toward the mesh. With a field of \Ea~=~11.3~kV/cm almost half of the charge is collected by the mesh.

The behavior of \Ime /\Itot, was fit to an exponential function:

\begin{equation}
\epsilon_{extr}(E_{\mathrm{A}}) = I_{\mathrm{mesh}}/I_{\mathrm{tot}} = A\cdot(1-e^{-E_{\mathrm{A}}/b})
\end{equation}

where $\epsilon_{extr}(E_{\mathrm{A}})$ is the fraction
of electrons exiting GEM~\#3 multiplication channels extracted and drifted to
the mesh as a function of \Ea\ while \Ig/\Itot\ is its complement to 1.
From a simultaneous fit, a value of A~=~0.57$\pm$0.02 was obtained indicating that, asymptotically, less than about 60\% of the charge can be collected on the mesh. This results is in good agreement with expectations from Garfield simulation found in literature \cite{bib:thesis, bib:Bonivento, bib:ieee_benci}.

\section{Evaluation of process mean free path}

From these measurements, the probability per unit path length \aex\ for an accelerated electron to create an excited neutral fragment CF$^*_3$, and thus a photon, can be evaluated. Let \nel\ be the total number of electrons and \ngem\ the total number of photons exiting from the last GEM. They are related by:

\begin{equation}
n_{\mathrm{GEM}} = n_{\mathrm{el}} \times \alpha_{\mathrm{GEM}} \times \Omega \times T
\label{eq:1}
\end{equation}

where \agem\ is the ratio between the number of photons and the number of electrons produced within the GEM, $\Omega$ is the fraction of the solid angle covered by the lens and $T$ is the optical transparency of the mesh plus the plastic foil beyond it.
The value of \ngem\ is the number of photons collected with \Ea~=~0~kV/cm measured to be 159~$\pm$~10 (see plot in Fig.\ref{fig:qIn}), while the value of \agem\ was published by our group in~\cite{bib:ieee_orange} and found to be about~$0.070\pm0.005$ in good agreement with other measurements~\cite{bib:Margato1}.

A similar relation is valid between \nel\ and the total amount of photons produced for electro-luminescence in the gap below the GEM (\nm(E$_{\mathrm{A}}$)):

\begin{equation}
n_{\mathrm{exc}}(E_{\mathrm{A}}) =  n_{\mathrm{el}} \times \epsilon_{extr}(E_{\mathrm{A}}) \times \alpha_{\mathrm{exc}}(E_{\mathrm{A}}) \times \Delta_z \times \Omega \times T 
\label{eq:2}
\end{equation}

where \nm\ can be measured as the increase in the number of collected photons with respect to \Ea~=~0 kV/cm. 
In particular $n_{\mathrm{exc}}$(11.3~kV/cm)~=~408~$\pm$~20.

By combining the equations \ref{eq:1} and \ref{eq:2}, the value of \aex\ can be obtained as:

\begin{equation}
\alpha_{\mathrm{exc}}(E_{\mathrm{A}}) = \alpha_{\mathrm{GEM}} \times \frac{1}{\epsilon_{extr}(E_{\mathrm{A}})} \times \frac{1}{\Delta_z} \times \frac{n_{\mathrm{exc}}(E_{\mathrm{A}})}{n_{\mathrm{GEM}}} = 1.2~\pm~0.1~{\mathrm{cm}} ^{-1}
\end{equation}

This value is in good agreement with the expectation of a mean free path of 1.0-2.0 centimetres for an electron to produce dissociation of CF$_4$ into neutral fragments as evaluated by \cite{bib:Fraga_beaune} and reported in Sect. \ref{sec:he-cf4}.

\section{Conclusion}

The possibility of improving the light signal in an Optical Readout triple-GEM structure by exploiting electro-luminescence in the gas outside the multiplication channels was studied.

A metal mesh was added 3 mm beyond the last GEM, to create an electric field to accelerate electrons and make them induce photon production by the gas molecules.
With this experimental set-up, an increase in the signal of about a factor 3 was obtained in stable working conditions by applying a 3.4~kV  voltage difference between the GEM and the mesh (i.e. an electric field of about 11.3~kV/cm).

Because of the very fast increase in the electro-luminescence cross-section as a function of the electric field strength, 
reaching slightly higher voltages with a better isolated setup would allow important increases of the scintillation yield. 

This technique can therefore easily lead to an interesting improvement in terms of light 
signal in Optically Readout GEMs without demanding for large electron gains.

\section*{Acknowledgment}
This work was supported by the European Research Council (ERC) under the European Union's Horizon 2020 research and innovation program (grant agreement No 818744).

\bibliographystyle{JHEP}
\bibliography{el}{}


\end{document}